\documentclass{elsart}

\usepackage{amssymb}

\begin{document}

\begin{frontmatter}



\title{Polarons and Solitons in Jahn-Teller Systems}


\author{Dennis P.~Clougherty}

\address{Department of Physics,
University of Vermont,
Burlington, VT 05405-0125 
USA}

\ead[url]{physics.uvm.edu/dpc/}

\begin{abstract}
Using a semiclassical continuum model of an electron in a deformable molecular crystal, some properties of multicomponent generalizations of the polaron--``vector polarons''-- are elucidated.  Analytical solutions for the case of two electronic bands coupled to two vibrational modes are given in detail.  Within the model considered, the vector polaron can be classified by its wavefunction into several types and can have features that include: (1) a spatial variation in the electronic and vibrational character, and (2) low-energy internal degrees of freedom.  For the case of electronic and vibrational degeneracy, local Jahn-Teller interactions can also lead to a novel spatiotemporal soliton, a long-lived excited state of the many-electron system stabilized by the conservation law resulting from degeneracy.  

\end{abstract}

\end{frontmatter}

\section{Introduction}
\label{intro}

Many of the complex solids currently under intensive theoretical and experimental study have common attributes.  High T$_c$ superconductors, fullerene-based materials, and colossal magnetoresistive manganites all feature electronic and vibrational degeneracies that are a consequence of the spatial symmetries of the crystal structure.  These systems typically have multiple electronic and structural phases competing at low temperatures.  And there is ample evidence \cite{dagotto05} that the electron-phonon interaction and the resulting polaron quasiparticles play an important role in shaping many of the physical properties of these materials.  

These observations serve as motivation for the theoretical study of the effects of generalized electron-phonon coupling in multiband systems.  Using a semiclassical continuum model of an electron in a deformable molecular crystal, the effects of electronic and vibrational degeneracy (and near degeneracy) on polaron structure have been investigated.  With the inclusion of local Jahn-Teller couplings within the degenerate manifolds, a multicomponent polaron emerges as a solution of the semiclassical equations of motion  \cite{dpc04}.

Experimentally the richer vibrational and electronic structure of this ``vector polaron'' might well be probed using STM to image the charge density in the vicinity of a vector polaron trapped at a defect.
This would be similar to the STM studies of the surface of an anisotropically gapped superconductor with impurities \cite{flatte}, where the order parameter symmetry has a great effect in shaping spatial features of the local electron density of states in the vicinity of an impurity, and this can be seen through measurements of the local conductivity.


 \section{The Model}
 \label{model}
 
 Consider a molecular solid with two orbitals per molecule: $|\psi_1\rangle$ and $|\psi_2\rangle$.  The molecule is also taken to have two vibrational modes: a breathing mode with amplitude $\phi_0$ and a symmetry-breaking mode with amplitude $\phi_1$.  The orbitals are taken to transform under elements of the molecular point group so that the symmetry-breaking mode couples to the difference in orbital densities, a local $E\otimes\beta$ Jahn-Teller term.  
 
 The continuum Hamiltonian for this system is taken to be
  \begin{equation}
  H= H_{el}+H_{ph}+H_{JT}
  \label{ham}
  \end{equation}
 with 
   \begin{equation}
 H_{el}=\int d^d x \Psi^\dagger  \left(-\frac{1}{2}\sigma_0\nabla^2+\Delta\sigma_3\right)  \Psi
 \end{equation}
  \begin{equation}
 H_{ph}=\int d^dx (\kappa_0 \phi_0^2/2+\kappa_1 \phi_1^2/2)
 \label{elec}
  \end{equation}
and 
   \begin{equation}
 H_{JT}=-\int d^dx \Psi^\dagger \left(g_0\phi_0\sigma_0+g_1\phi_1\sigma_3\right)  \Psi
 \label{e-jt}
  \end{equation}

 where $\Psi(x)$ is a two-component field operator.  The upper component destroys an electron from the first orbital at $x$, while the lower component destroys an electron from the second orbital at $x$.
 In the absence of distortions, the band constructed from $|\psi_1\rangle$ states is offset in energy by $2\Delta$ from the band constructed from  $|\psi_2\rangle$ states. The limit $\Delta\to 0$ corresponds to doubly degenerate bands.  
 
 A continuum approximation to the local $E\otimes\beta$ model studied by H\"ock {\it et al.} \cite{hock} is recovered in the limit $g_0\to 0$, while the continuum Holstein molecular crystal model \cite{holstein} is obtained in the limit $g_1\to 0$.  Hence, the model of Eq.~\ref{ham} combines the Holstein model and the $E\otimes\beta$ model.

 \section{Bogoliubov-de Gennes Equations}

 The Bogoliubov-de Gennes (BdG) equations for this model can be found by expanding the field operator in terms of a new real-space basis.  Envelope functions $\{u_n, v_n\}_{n=1}^\infty$ that modulate the molecular orbitals are introduced, and the field operator becomes
 \begin{equation}
  \Psi=\sum_n\left(\matrix{u_n\cr
  v_n}\right)c_n
 \end{equation}
 where $c_n$ destroys an electron with amplitudes $u_n(x)$ and $v_n(x)$ for the two molecular orbitals.  A necessary condition (unitarity) on the amplitudes is then 
 \begin{equation}
 \int d^d x (u_n^*u_{n'}+v^*_n v_{n'})=\delta_{nn'}
 \end{equation}
 The BdG equations for the envelope functions are then found to be
  \begin{eqnarray}
\left(-\frac{1}{2}\nabla^2+\Delta -g_0\phi_0 -g_1\phi_1\right)u_n=\epsilon_n u_n\nonumber\\
\left(-\frac{1}{2}\nabla^2-\Delta -g_0\phi_0 +g_1\phi_1\right)v_n=\epsilon_n v_n
  \label{bdg}
 \end{eqnarray}
 
 From energy minimization, the distortions must satisfy the following self-consistent equations
  \begin{eqnarray}
  \phi_0&={g_0\over\kappa_0}\sum_n (|u_n|^2+|v_n|^2)\nonumber\\
 \phi_1&={g_1\over\kappa_1}\sum_n (|u_n|^2-|v_n|^2)
  \label{self}
  \end{eqnarray}
 where the sums are over all occupied states.
 
 For a system with a single electron, only one term contributes to the sum.    Eqs.~\ref{self} can be substituted into the BdG equations to obtain the two coupled nonlinear Schr\"odinger equations for the envelope functions $u, v$ previously obtained in Ref.~\cite{dpc04}
\begin{eqnarray}
\left(-\frac{1}{2}\nabla^2-\nu |u|^2-\eta |v|^2\right)u=(\epsilon-\Delta) u\nonumber\\
\left(-\frac{1}{2}\nabla^2-\eta |u|^2-\nu |v|^2\right)v=(\epsilon+{\Delta}) v
\label{nlse}
\end{eqnarray}
where $\nu\equiv {g_0^2/\kappa_0}+{g_1^2/\kappa_1}$ and $\eta\equiv {g_0^2/\kappa_0}-{g_1^2/\kappa_1}$.

The single electron might also be thought of as occupying a conduction band or partially occupying a set of conduction bands.  Provided the polarization effects from the filled valence band are small, the ``frozen'' valence band approximation can be made, and the identical BdG equations result.

The spatial coordinates and the envelope functions can be scaled
\begin{eqnarray}
\xi=\sqrt{2|\epsilon-\Delta|}x\nonumber\\
\left(\matrix{r_1\cr r_2}\right)=\sqrt{\nu\over|\epsilon-\Delta|} \left(\matrix{u\cr
  v}\right)
\end{eqnarray}
to yield the following equations
\begin{eqnarray}
\left(\nabla^2-1+r_1^2+\beta r_2^2\right)r_1=0\nonumber\\
\left(\nabla^2-\omega^2+r_2^2+\beta r_1^2\right)r_2=0
\label{nlse2}
\end{eqnarray}
where $\omega^2\equiv{(|\epsilon|-\Delta)/(|\epsilon|+\Delta)}$ and $\beta=\eta/\nu$.

Eqs.~\ref{nlse2} have been previously studied in many different contexts such as nonlinear optics (propagation of light in birefringent optical fibers \cite{malomed}) and atomic physics (spinor Bose-Einstein condensation \cite{ho}) where the resulting Ginzburg-Landau equations have the same form.  Although the mathematical study of systems of coupled nonlinear Schr\"odinger equations is still in its infancy, a set of known solutions in 1D and 2D to these equations that describe the components of the vector polaron wavefunction will now be summarized.  

 \section{Vector Polarons in 1D}
 
 For one spatial dimension, three regimes can be identified where analytic solutions to Eq.~\ref{nlse2} that are stable and localized are known to exist.  In general, it is required that $\epsilon <0$ and $|\epsilon|<\Delta$.  $\eta$ and $\nu$ are restricted here to be non-negative.   Thus, $0\le\beta\le 1$.
  
\subsection{Manakov type}
 The case of $\beta=1$ and $\omega=1$  is special, as it corresponds to an integrable system.  Manakov \cite{manakov} studied this system of equations in another context.  The symmetry of these equations gives rise to an infinite set of degenerate solutions that can be labeled by a continuous parameter $0\le \theta<2\pi$.  
  \begin{eqnarray}
r_1(\xi)=\sqrt{2} \cos\theta\  {\rm sech}\  \xi\nonumber\\
r_2(\xi)=\sqrt{2} \sin\theta\  {\rm sech}\ \xi
\end{eqnarray}

$\omega=1$  corresponds to exact degeneracy $(\Delta=0)$, while $\beta=1$ results from no coupling to the symmetry-breaking mode $(g_1\to 0)$.  Consequently, the energy of the state is independent of orbital density differences.
Thus the Manakov solution contains a zero-energy internal degree of freedom, corresponding to periodic changes in the local orbital mixing.  

 \subsection{Equal Amplitude type}
 One situation that is exactly soluble is the ``equal amplitude'' case where $r_1(\xi)=r_2(\xi)$.  For stable, localized solutions, the parameters must be further restricted so that $\omega=1$.  In this case, Eqs.~\ref{nlse2} reduce to 
 \begin{equation} 
 \left({d^2\over d\xi^2}-1+(1+\beta) r^2\right)r=0
 \end{equation}
 The solution to this scalar nonlinear Schr\"odinger equation is the well-known bell soliton
  \begin{equation} 
r(\xi)=\sqrt{2\over 1+\beta}\ {\rm sech}\ \xi
 \end{equation}
The continuous symmetry present in the Manakov case is broken for $\beta\ne 1$.  However, rotations of the orbital mixing still serve as a low-energy degree of freedom for $\beta$ close to 1.
  
 \subsection{Wave-Daughter Wave type}
 For $\omega< 1$, corresponding to non-degeneracy of the orbitals, it is possible to have one envelope function dominate throughout the core of the localized wavefunction.  In this wave-daughter wave approximation, the amplitude of the daughter wave $r_2$ is much smaller than that of the primary wave $r_1$, leading to the following approximate set of equations
 \begin{eqnarray}
\left({d^2\over d\xi^2}-1+r_1^2\right)r_1\approx 0\nonumber\\
\left({d^2\over d\xi^2}-\omega^2+\beta r_1^2\right)r_2\approx 0
\end{eqnarray}

 The top equation is recognized as the scalar nonlinear Schr\"odinger equation with a solution
 \begin{equation} 
r_1(\xi)=\sqrt{2}\  {\rm sech}\ \xi
\label{primary}
 \end{equation}
The primary wave provides an attractive potential for the daughter wave, and the P\"oschl-Teller equation for the daughter wave results
 \begin{equation} 
\left({d^2\over d\xi^2}-\omega^2+2\beta {\rm sech}^2\xi\right)r_2=0
 \end{equation}

This is known \cite{landau} to have a single bound state 
for $0<\beta\le 1$, 
with
\begin{equation} 
r_2(\xi)=\alpha\  {\rm sech}^\omega\ \xi
 \end{equation}
and an eigenvalue constraint that  $\omega =  ( \sqrt{1+8 \beta} -1)/2$.

Within these cases amenable to analytical solutions, it is seen that (1) vector polaron states can support low-energy internal degrees of freedom, corresponding to pseudo-rotations of the electron amplitude in the orbital space; and (2) vector polaron states can have spatial variations in the orbital mixing and in the character of the distortion, as illustrated in the wave-daughter wave type.

 \section{Vector Polarons in 2D}

Vortex vector solitary wave solutions are known to exist in 2D.  Consider solutions to Eqs.~\ref{nlse2} of the form $r_i(\rho,\phi)=\psi_i(\rho)\exp(im_i\phi)$.  The following coupled equations result
\begin{eqnarray}
\left(\Delta_\rho-{m_1^2\over\rho^2}-1+\psi_1^2+\beta \psi_2^2\right)\psi_1=0\nonumber\\
\left(\Delta_\rho-{m_2^2\over\rho^2}-\omega^2+\psi_2^2+\beta \psi_1^2\right)\psi_2=0
\label{radial}
\end{eqnarray}
where $\Delta_\rho\equiv {1\over\rho}{d\over d\rho}\left(\rho{d\over d\rho}\right)$.   Vortex solutions to Eq.~\ref{radial} may be labeled by the topological charges $|m_1,m_2\rangle$.  

The simplest case is the $|0,0\rangle$ solution, where both envelope functions are radially symmetric.  Using an analytical variational method, the existence of a localized $|0,0\rangle$ solution has been shown \cite{malmberg}.  Unfortunately this solution has also been shown to be linearly unstable, as are all of these vortex solitons.  Related multipole vector solitons \cite{kivshar} are also linearly unstable in this simplified model.

However, similar solutions that are linearly stable have been found for slightly modified systems of equations that include either saturable nonlinearities \cite{malmberg} or nonlocal nonlinearities \cite{kartashov}.  Saturable nonlinearities would physically correspond to the addition of proper vibrational anharmonicity of higher orders in this model, while nonlocal nonlinearity would correspond to gradient terms in the elastic energy.  Consequently there are indications that semiclassical multicomponent polaron states of the vortex and multipole varieties stably exist in generalized models of 2D systems.

 \section{Jahn-Teller Soliton}
Another related continuum model that employs a local Jahn-Teller coupling can also lead to a stable  soliton, a state involving a lump of excess electronic charge trapped by a time-dependent pseudo-rotation of the elastic field \cite{dpc06}. These solitons fall in the class  of solutions known as non-topological Q-balls \cite{coleman} that have been invoked in models of cosmological phase transitions in the early Universe \cite{gleiser}.

Consider a continuum model with three electronic bands linearly coupled to two degenerate elastic modes.  Take the case of where the electronic bands transforms as $x$ and $y$ and a trivial orbital singlet; the two elastic modes transform as $x$ and $y$. The system may be described by the following  Hamiltonian 
\begin{equation}
H=H_e+H_{ph}+H_{JT}
\label{soliton}
\end{equation}
where
\begin{eqnarray}
H_e&=\sum_{m=0,\pm 1} \int d^3x (-\frac{1}{2}\psi_m^\dagger \nabla^2 \psi_m+W\delta_{|m|, 1} \psi_m^\dagger  \psi_m)\\
H_{ph}&=\sum_{m=\pm 1}\int d^3x (|\partial_t \phi_m |^2+U(|\phi_m|^2)\ \ \ \  \ \ \ \  \ \ \ \ \ \ \ \ \\
H_{JT}&=-g\sum_{m=\pm 1} \int d^3x (\phi_{-m} \psi_m^\dagger  \psi_0 + H.c.)\ \ \ \  \ \ \ \  \ \ \ \  
\end{eqnarray}
where $W$ is the energy splitting between the doublet bands and the singlet, and $m$ labels the states in the axial angular momentum basis. 

Such a system has a continuous symmetry whose origin lies in the electronic and vibrational degeneracy.  By virtue of Noether's theorem, a conserved charge $Q$ results from this continuous symmetry provided it remains unbroken.  The constraint of fixed $Q$ provides stability to the soliton state.

For large $Q$, a variational ansatz called the thin wall approximation \cite{coleman} has been found to give good results in calculating the soliton energy.   The amplitude of the elastic deformation is taken to be constant over a finite region of space with volume $V$, while vanishing outside that region.  This deformation itself, however, is not constant; rather, it undergoes pseudorotations in the two-fold degenerate space.  It is straightforward to show that the Noether charge $Q$ is linearly proportional to the pseudorotation frequency $\omega$ and to the square of the amplitude of distortion \cite{dpc06}
 \begin{equation}
Q\approx\omega\xi^2 V
\label{q}
\end{equation}

Thus, for sufficiently large band splitting $W$, the energy of this thin-walled soliton  is 
\begin{eqnarray}
E=&(K_{ph}+U -E_{JT})V\ \ \ \ \ \ \ \ \ \ \ \ \ \nonumber\\
\approx&\frac{1}{2}\omega^2\xi^2V+U(\xi^2)V-{g^2 n_0\over W}\xi^2 V\nonumber\\
\approx&{Q^2\over 2 \xi^2V}+\tilde U(\xi^2)V\ \ \ \ \ \ \ \ \ \ \ \ \ \ \ \ \ 
\label{energy}
\end{eqnarray}
where $K_{ph}$ and $U$ are the kinetic and potential energy density of the time-dependent distortion, $E_{JT}$ is the energy density from the electron-phonon interaction, $n_0$ is the electron density, and
$\tilde U\equiv U(\xi^2)-{g^2 n_0\over W}\xi^2$.
The last step follows from using Eq.~\ref{q}  to eliminate $\omega$ in favor of $Q$ and $\xi$.

As Coleman \cite{coleman} showed, the resulting energy can be minimized subject to fixed Q to find the soliton energy and its size.  
\begin{eqnarray}
E=&Q\sqrt{2\tilde U\over\xi^2}\\
V=&{Q\over\sqrt{2\tilde U \xi^2}}
\label{energy2}
\end{eqnarray}
Thus provided that the minimum value of $2{\tilde U/\xi^2}$ occurs at $\xi\ne 0$, this Jahn-Teller soliton cannot decay by emitting phonons that carry away the charge Q, since the energy per charge is lower for the soliton.  This criterion sets the requirements on the amount and type of anharmonicity that are necessary to stabilize the soliton.

In summary, it is noted that many of the solitons, both topological and non-topological, that are found in multicomponent Ginzburg-Landau theories can be used to describe vector polarons within the semiclassical theory outlined here. Experimental STM studies of these states trapped at defects might be used to reveal the non-trivial spatial structure of these polarons.  Vector polarons, together with the many-electron Q-ball-like excited state, serve as examples that systems containing electronic and vibrational degeneracy can support novel states through local Jahn-Teller interactions.  

 \section{Acknowledgments}
 This work was partially supported by the National Science Foundation through a grant for the Institute for Theoretical Atomic, Molecular and Optical Physics at Harvard University and the Smithsonian Astrophysical Observatory.
 \vfil\eject

\end{document}